\documentclass[aps,prb,twocolumn,superscriptaddress]{revtex4-2}
\usepackage{bm}
\usepackage{ae}
\usepackage[T1]{fontenc}
\usepackage[ansinew]{inputenc}
\usepackage{amsmath}
\usepackage{amssymb}
\usepackage{graphicx}
\usepackage{color}
\usepackage[colorlinks]{hyperref}
\usepackage{epstopdf}
\input{epsf}

\begin{document}

\title{Nanomechanical manipulation of superconducting charge-qubit quantum networks}

\author{D. Radi\'{c}*}
\affiliation{Department of Physics, Faculty of Science, University of Zagreb, Bijeni\v{c}ka 32, Zagreb 10000, Croatia}

\author{L.Y. Gorelik}
\affiliation{Department of Physics, Chalmers University of Technology, SE-412 96 G{\"o}teborg, Sweden}

\author{S.I. Kulinich}
\affiliation{B. Verkin Institute for Low Temperature Physics and Engineering of the National Academy of Sciences of Ukraine, 47 Prospekt Nauky, Kharkiv 61103, Ukraine}

\author{R.I. Shekhter}
\affiliation{Department of Physics, University of Gothenburg, SE-412 96 G\"oteborg, Sweden}

%\date{\today}
%\pacs{}

\begin{abstract}

We suggest a nanoelectromechanical setup and corresponding time-protocol for controlling parameters in order to demonstrate nanomechanical manipulation of superconducting charge-qubit quantum network.  
We illustrate it on an example reflecting important task for quantum information processing - transmission of quantum information between two charge-qubits facilitated by nanomechanics.
The setup is based on terminals utilizing the AC Josephson effect between bias voltage-controlled bulk superconductors and mechanically vibrating mesoscopic superconducting grain in the regime of the Cooper pair box, controlled by the gate voltage. The described manipulation of quantum network is achieved by transduction of quantum information between charge-qubits and intentionally built nanomechanical coherent states, which facilitate its transmission between qubits. This performance is achieved using quantum entanglement between electrical and mechanical states. 
\end{abstract}

\maketitle

\section{Introduction}

Quantum networks are main tools to provide transfer of quantum information (QI) \cite{Nielsen} during its processing, thus constituting an essential element of quantum computing and quantum communication systems. In the wider sense, quantum networks provide connection between distinct quantum processors or even computers in the analogous way to what classical networks do with classical processors or computers \cite{Kimble}. In the narrower sense, a set of constituents of a quantum processor, qubits, connected in the way to be able to transmit the QI between themselves to form special kinds of entangled quantum states, represent a quantum network as well. The latter is a subject of this paper.
Qubit \cite{Schumacher} is a quantum system with two states where the QI is stored in their superposition \cite{Nielsen}. It is a basic "container" of the QI in the quantum network that we study.
The very implementation of qubit has been a subject of extensive research, covering number of fields in physics such as optics, atomic physics or solid-state physics, as well as their combinations \cite{QuantumProcessing,Girvin,Devoret,Mirhosseini,Leibfried}, in an attempt to find an optimum between cons and pros of each implementation. In this paper we are focused to the nanoelectromechanical (NEM) implementation in which we utilize coherent interplay of qubit states and nanomechanical excitations \cite{Schneider,Hann,Chu}. The main purpose of this interplay is to achieve transduction of the QI between qubit and nanomechanical subsystems.
One brilliant demonstration of such transduction was recently realised by coupling of a superconducting qubit circuit to a mechanical surface resonator \cite{Cleland1,Cleland2} where the individual phonons were controlled and detected by a superconducting qubit circuit, enabling a coherent generation and measurement of a non-classical superposition of the zero- and one-phonon itinerant Fock states. This type of control enables a phonon-mediated quantum state transfer and establishing quantum entanglement of the remote qubits.  Utilisation of nanomechanics appears as a natural choice to deal with transduction, transmission, or storing of the QI due to amazingly high quality factors achievable nowadays \cite{Laird,Tao,Bereyhi}. 
Quantum network, studied in this paper, adopts the NEM implementation of terminals based on the AC Josephson effect. It comprises voltage-biased superconducting leads, with a superconducting mesoscopic grains, attached to a single nanomechanical oscillator, placed amid. States in each grain are controlled by the gate voltage in the way to be in regime of the Cooper pair box (CPB). Those are charge-qubit in our setup \cite{Bouchiat,Nakamura,Robert,Lehnert}. 
In our first publications on this topic \cite{npjDR,Bahrova} we showed that dynamics of Josephson tunneling between the CPB and the leads, coupled to mechanical oscillations, resulted in formation of nanomechanical coherent states entangled with states of a charge-qubit if Josephson and mechanical frequencies are in resonance. Furthermore, applying the bias voltage manipulation protocol, we demonstrated an onset of nanomechanical cat-states consisting of coherent states.
In the later publication we showed that the QI, encoded into qubit states, can be transduced into the pure nanomechanical cat-state and back, providing the time-protocol of manipulating the bias and gate voltages for that functionality \cite{PhysB_DR}. Here, the entanglement itself appears as a powerful resource for this particular type of functionality, but it also appears as such in the field of quantum communication and computation in general \cite{Horodecki}.\\  
%
%
%%%%%%%%%%%%%%%%%%%%%%%%%%
\begin{figure*}
\centerline{\includegraphics[width=2\columnwidth]{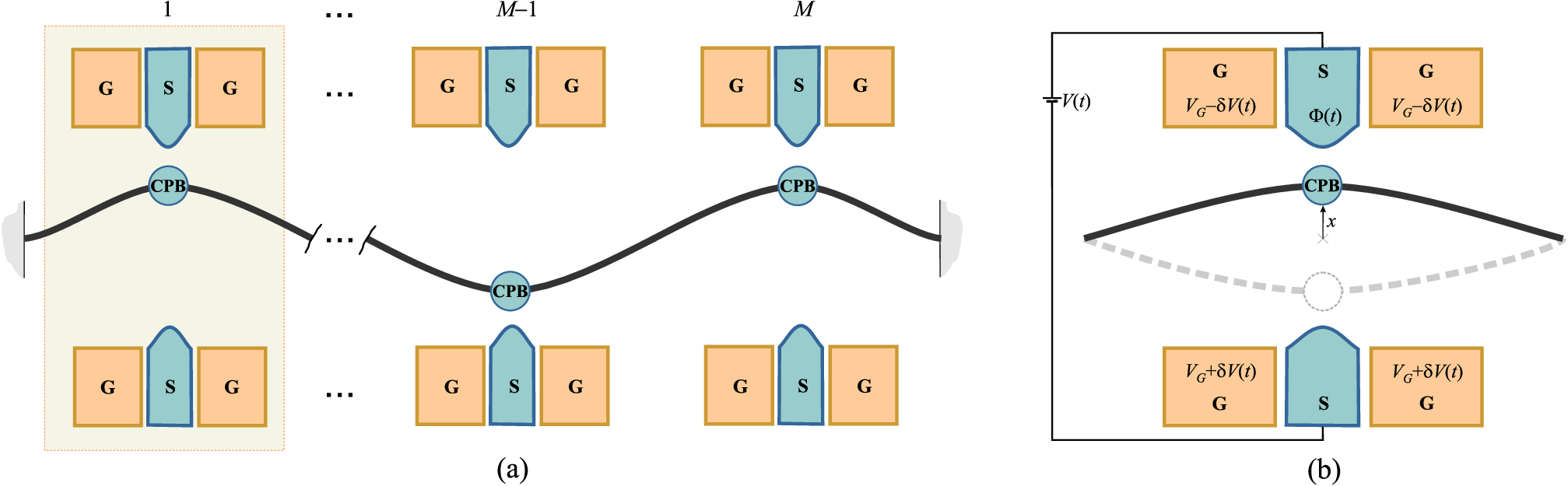}}
\caption{Schematic illustration of the NEM implementation of the $M$-terminal quantum network. (a) Quantum network containing $M$ qubits - Cooper pair boxes (CPB) attached to the same suspended beam performing mechanical vibrations in the $M$-th flexural mode. CPBs are placed at positions where elongations of mechanical vibrations are maximal. For each CPB there is a superconducting Josephson junction (S) and gate electrodes (G), which together make one terminal of the network (shaded). Each terminal is electrically wired independently. (b) One terminal of the network: The in-plane harmonic vibrations, characterized by displacement $x$, appear between two superconducting electrodes (S) symmetrically biased by a constant voltage $V$ used to create a superconducting phase difference $\Phi(t)$ between S-electrodes. Electrostatic gate electrodes (G) usage is twofold: $V_G$ is used to bring the states of the quantum dot with zero and one excess Cooper pair into the CPB regime, while $\delta V(t)$ is used in the time-protocol to provide a finite approximately homogeneous electric field across the central region of the junction (cross) without changing the potential in the origin (due to symmetric configuration).}
\label{FigSchematic}
\end{figure*}
%%%%%%%%%%%%%%%%%%%%%%%%%%%%
%
%
The quantum network that we propose (see Fig. \ref{FigSchematic}) consists of $M$ such CPBs, i.e. of $M$ qubits, attached to the same suspended beam that performs harmonic mechanical oscillations in the $M$-th flexural mode. CPBs are placed at positions where maxima of elongation take place. For each CPB there exists an independently wired controlling terminal consisting of superconducting Josephson junction and electrostatic gate electrodes, as described above. 
Using the property of coupled resonant dynamics of charge-qubits and nanomechanics, entanglement, as well as transduction of the QI via nanomechanical cat-states, one can achieve different functionalities of such a network finding  specific time-protocols for controlling parameters for every particular function. Just two examples of such functionalities, which are important tasks in the QI processing, would be: (1) for the network with $M=2$, one can demonstrate transmission of the QI between two qubits, i.e. $\alpha \mid \uparrow \uparrow \rangle + \beta \mid \downarrow \uparrow \rangle \rightarrow \alpha \mid \uparrow \uparrow \rangle + \beta \mid \uparrow \downarrow \rangle$; (2) for the network with $M=3$, one can create a special, entangled 3-qubit state, with the initial QI encoded inside it, i.e. $\alpha \mid \uparrow \uparrow \uparrow \rangle + \beta \mid \downarrow \uparrow \uparrow \rangle \rightarrow \alpha \mid \uparrow \uparrow \uparrow \rangle + \beta \mid \downarrow \downarrow \downarrow \rangle$, which has a special role in the theory of quantum error correction, so-called {\it the bit flip code} \cite{Peres}. Here $\mid \uparrow \rangle$, $\mid \downarrow \rangle$ denote two possible states of each qubit, while complex amplitudes $\alpha$, $\beta$ define the quantum superposition of states, under the unitarity condition $\vert \alpha\vert^2+\vert\beta\vert^2 =1$, containing the QI. In this paper we elaborate the first example on 2-terminal quantum network, carrying out a detailed calculations of the operating time-protocol required to achieve transmission of the QI from one qubit to another. 

The paper is organized in the following way: after the Introduction, in Section II we present the setup and model Hamiltonian of the system; in Section III we present the  time-evolution operators of the system, based on Hamiltonian and time-protocols of operating the controlling parameters, necessary for the desired functionality; in Section IV we elaborate the 2-qubit network and time-protocol for transmission of the QI between qubits; section Conclusions contains the concluding remarks and discussion.

\section{The Model}

Schematic picture and description of the $M$-qubit quantum network is shown in Fig. \ref{FigSchematic}.
Symmetrically placed gate electrodes, within each terminal $l=1,2,\ldots,M$, have two-fold function: (1) By the particular choice of potential $V_G$ on gate electrodes, the mesoscopic superconducting grain is set into regime of the Cooper pair box, i.e. the effective two-level system of degenerate states with zero and one excess Cooper pair, denoted by $\mid \downarrow \rangle$ and $\mid \uparrow \rangle$ respectively. Those we call "the charge states" or "the CPB qubit states".
(2) Applying a constant additional voltage $\delta V^{(l)}$ over the gate electrodes, we can create an approximately homogeneous electric field $\mathcal{E}^{(l)}$ along the central part of the junction $l$ where the CPB$^{(l)}$ moves, with a zero-value of the corresponding electrostatic potential in the middle (i.e. approximately preserving the degeneracy of the CPB states). 
While $V_G$ is kept constant, switching the $\delta V^{(l)}$ on and off is an important part of the time-protocols of transduction of the QI and achieving the desired functionality as will be shown in the paper.\\
 
The superconducting part of the junction (S-leads and the CPB oscillating between them) within each terminal operates in the regime of the AC Josephson effect, i.e. the superconducting electrodes are biased by a constant symmetric bias voltages $V^{(l)}(t)$, providing a superconducting phase difference $\Phi^{(l)}(t)$ dependent on time $t$, where $\Phi(t)=\mathrm{sgn}(V)\,\Omega t$ and $\Omega = 2e|V|/\hbar$ is the Josephson frequency assumed to be the same for all junctions. Some constant initial phase at $t=0$ is for simplicity taken to be zero. Cooper pairs tunnel between the S-electrodes and the CPB, which is attached to a suspended nanobeam performing mechanical vibrations at frequency $\omega$, thus making the tunneling essentially position-dependent. Neglecting a geometric asymmetry of the junction, we expand the  Josephson coupling in terms of a small parameter $\varepsilon \equiv x_0 / x_{\text{tun}} \ll 1$, where $x_0=\sqrt{h/m\omega}$ is the amplitude of zero-mode oscillations ($m$ is mass of the oscillator) and $x_{\text{tun}}$ is the tunneling length. In this model we assume symmetric junctions fabricated equally, having the Josephson energy parametar equal for all junctions, i.e. $E_J^{(l)}=E_J$, $\forall l$. \\

Finally, as all CPBs are attached to the same beam (e.g. a carbon nanotube), performing harmonic mechanical oscillations in the $M$-th flexural mode with frequency $\omega$, there is a single harmonic oscillator coupled to all CPBs through the position-dependent terms of Josephson coupling.\\ 

Taking it all into account, with coupling approximately linear in displacement of the CPBs, we write the time-dependent model Hamiltonian of the network in the form
% 
%%%%%%%%%%%%%%%%%%%%%%%%%% 
\begin{eqnarray}\label{Hamiltonian}
H(t)= \hbar\omega b^\dag b
&+& \sum_{l=1}^M \left\lbrace E_J\sigma_x^{(l)}\cos \Phi^{(l)}(t) \right. \nonumber\\
&-& (-1)^l \left. \varepsilon E_J \hat X \sigma_y^{(l)} \sin \Phi^{(l)} (t)\right. \nonumber\\ 
&-& (-1)^l \left. E^{(l)}(t) \hat X \sigma_z^{(l)} \right\rbrace,
\end{eqnarray}
%%%%%%%%%%%%%%%%%%%%%%%%%%
%
where index $l$ denotes terminal within the network, also marking the corresponding operators. Factor $(-1)^l$ accounts for the sign of the displacement operator having in mind that neighbouring CPBs move in the opposite directions.
For each terminal, $E^{(l)}(t)=2|e|\mathcal{E}^{(l)}(t)x_0$ is the (eletrostatic) energy related to electric field $\mathcal{E}^{(l)}$ generated by $\delta V^{(l)}(t)$, $\sigma_{x,y,z}^{(l)}$ are the Pauli matrices operating in the  $2 \times 2$ Hilbert subspace of the CPB$^{(l)}$. Operator $b^{\dagger}$ is the phonon creation operator, acting in the mechanical subspace, as well as the coordinate operator $\hat X = (b+b^\dag)/\sqrt{2}$.
%Hamiltonian (\ref{Hamiltonian}) is written in so-called charge representation where the qubit states are $\mid \downarrow \rangle = (0\,\,1)^{\text{T}}$ and $\mid \uparrow \rangle=(1\,\,0)^{\text{T}}$.

\begin{figure}
\centerline{\includegraphics[width=\columnwidth]{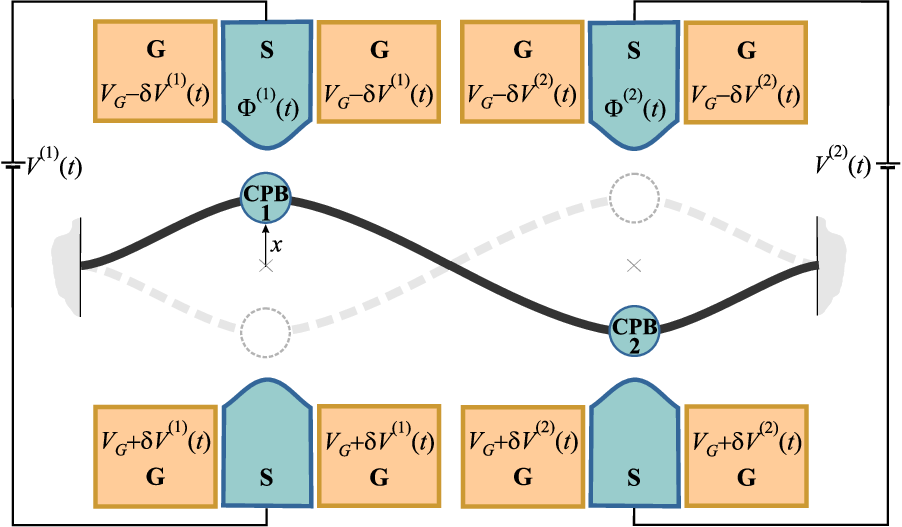}}
\caption{Schematic illustration of the 2-qubit quantum network with two terminals wired independently in order to control parameters: bias voltages $V^{(1,2)}(t)$ and gate voltages $\delta V^{(1,2)}(t)$ at terminals 1 and 2.}
\label{Fig2qubit}
\end{figure}

In this paper we focus in detail to the example of 2-qubit network ($M=2$) in which we demonstrate transmission of the QI between qubits. The corresponding schematic setup is shown in Fig. \ref{Fig2qubit}. 
%
%
%%%%%%%%%%%%%%%%%%%%%%%%%%
%%%%%%%%%%%%%%%%%%%%%%%%%%%%
%
%

As we have shown in Refs. \cite{npjDR,PhysB_DR}, the coherent dynamics in the setup with one qubit, under the resonant condition $\omega=\Omega$, gives onset to nanomechanical coherent states $\vert Z(t) \rangle$. If we limit our consideration to the case when the term with electric field (last term in the Hamiltonian (\ref{Hamiltonian})) acts on the well-developed coherent state, we can approximate the mechanical displacement operator $\hat{x}=x_0 \hat X$ with its classical (mean) value $x(t)=A(t) \cos \omega t$, analogous to the mean-field approximation. For the nanooscillator with resonant driving in our consideration, $x(t)$ corresponds to an eigenvalue of the coherent state $\vert Z(t) \rangle$ with an amplitude $A(t)$ linearly growing in time. Corrections are of the order of $x_0$ which is much smaller than the amplitude of resonant oscillations. We keep track of the coherent state corresponding to $x$ by writing $x \rightarrow x_Z$. Furthermore, in our consideration we limit the action of that term on time-evolution of the system along the time interval when coupling of the coherent Josephson dynamics to harmonic oscillator is absent, so the coherent states, once developed, perform only the unitary rotation in the phase space, i.e. the amplitude of oscillations does not grow in time, changing only the phase. Considering it all, we write the Hamiltonian of the 2-qubit network, valid under the above-stated conditions, in the form
% 
%%%%%%%%%%%%%%%%%%%%%%%%%% 
\begin{eqnarray}\label{Hamiltonian2CPB}
H(t)\approx \hbar\omega b^\dag b
&+& \sum_{l=1}^2 \left\lbrace E_J\sigma_x^{(l)}\cos \Phi^{(l)}(t) \right. \nonumber\\
&-& (-1)^l \left. \varepsilon E_J \hat X \sigma_y^{(l)} \sin \Phi^{(l)} (t)\right. \nonumber\\ 
&-& (-1)^l \left. E^{(l)}_Z (t)\, \sigma_z^{(l)} \right\rbrace,
\end{eqnarray}
%%%%%%%%%%%%%%%%%%%%%%%%%%
%
where $E^{(l)}_Z (t)=2|e|\mathcal{E}^{(l)}(t)x_Z(t)$ in which by the free index $l$ we denote that we can control electric field for each terminal independently (controlling the corresponding gate voltage).\\

\section{The time-evolution operators}

Since all both terminals in the network are independently wired, we can operate controlling parameters $V^{(1,2)}(t)$ and  $\delta V^{(1,2)}(t)$ at any given time $t$ for each terminal. In that sense, the time-evolution of the system can be controlled for each terminal. However, there is a single nanomechanical oscillator related to all terminals, meaning that whatever mechanical state is generated operating one terminal, it will affect all others if the coupling is switched on. This very property of the mechanical subsystem within the network will be the main QI transduction and entanglement tool to manipulate states in all qubits.\\

In an example of network manipulation to be presented in the paper, we start with prepared pure initial quantum state of the 2-qubit system at $t=0$,
% 
%%%%%%%%%%%%%%%%%%%%%%%%%% 
\begin{eqnarray}\label{Psi_0}
\Psi(t=0)=\left( \alpha \vert \mathbf{e}_y^+\mathbf{e}_x^- \rangle + \beta \vert \mathbf{e}_y^-\mathbf{e}_x^- \rangle \right)\vert 0 \rangle,
\end{eqnarray}
%%%%%%%%%%%%%%%%%%%%%%%%%%
%
where $\mathbf{e}_i^{\pm}$ are the eignevectors of $\sigma_i$ Pauli matrix corresponding to eigenvalues $\pm 1$, while the complex coefficients $\alpha$ and $\beta$, $\vert \alpha\vert^2+\vert\beta\vert^2 =1$, contain the QI. Position of the vector in the ket state $\vert \mathbf{e}^{\pm\,(1)}_i \mathbf{e}^{\pm\,(2)}_j \rangle$ denotes the terminal, i.e. the first refers to the state of qubit 1 (Q1), while the second one accounts for qubit 2 (Q2).
Initially, the harmonic oscillator is in the ground state with zero phonons $\vert 0 \rangle$. 
The initial wave function has the QI encoded in the pure state, the superposition of states of Q1 while Q2 is the same and can be factorized out.
In order to achieve the desired functionality of the network, i.e. to transmit the QI into pure state, superposition of the Q2 states with Q1 factorized out, we shall need three types of time-evolution operators that follow, based on Hamiltonian (\ref{Hamiltonian2CPB}) and time-protocol of manipulation of its terms.
The idea is to use independent activation of terminals during the time-evolution to provide coherent resonant dynamics between the Josephson tunneling junction and nanomechanical oscillator which "inflate" and "deflate" nanomechanical coherent states entangled with qubit states, for which first three terms in Hamitonian (\ref{Hamiltonian2CPB}), i.e. the "$\hbar\omega$"-, the "$E_J$"-, and the "$\varepsilon$"-term, are responsible.
The last, "$E_Z$"-term in (\ref{Hamiltonian2CPB}) operates when coherent states are well-developed and disconnected from qubit dynamics, serving to "rotate" the targeted qubit states to the desired one in order to achieve the pure state.
These are two main tools for transduction of the QI between qubit subspace and nanomechanical subspace. Carefully tailored protocol of operation provides, in that sense, transmission of the QI between different qubits using nanomechanics as a mediator.\\

The time-evolution operator $\hat U(t,t_0)$ of the wave function from time moment $t_0$ to $t$ is generally solution of the equation $i\hbar \tfrac{\partial}{\partial t}\hat U(t,t_0)=H(t) \hat U(t,t_0)$, with initial condition $\hat U(t_0,t_0)=\mathbf{1}$. It can be found by number of standard methods of quantum mechanics, depending on the properties of $H(t)$, the most popular among them being the interaction picture representation.
Three types of time-evolution operators, that we need to describe qubit - nanomenchanics evolution for each terminal of the network, are:\\

(I): The time-evolution operator
%
%%%%%%%%%%%%%%%%%%%%%%%%%%%%
\begin{eqnarray}\label{U_I}
\hat U_{\mathrm{I}}(NT)= e^{-i 2\pi N b^\dag b} e^{\pm i \lambda N \hat P \sigma_y},
\end{eqnarray}
%%%%%%%%%%%%%%%%%%%%%%%%%%%%
%
%
%%%%%%%%%%%%%%%%%%%%%%%%%%%%
%\begin{eqnarray}\label{U_I}
%\hat U_{\mathrm{I}}(NT)= e^{-i 2\pi N b^\dag b} e^{i \lambda N %\hat P \sigma_y^{(1)}} e^{-i \lambda N \hat P \sigma_y^{(2)}},
%\end{eqnarray}
%%%%%%%%%%%%%%%%%%%%%%%%%%%%
%
where $\lambda=\varepsilon\pi\tfrac{E_J}{\hbar\omega} (J_0(\tfrac{2E_J}{\hbar\omega})-J_2(\tfrac{2E_J}{\hbar\omega}))$, is given in terms of the Bessel function of the first kind $J_k$.
In this form, it is obtained for the interval of time equal to the integer number of oscillation periods $\Delta t = NT$, $T=2\pi/\omega$, from Hamiltonian (\ref{Hamiltonian2CPB}) in the absence of the last term, i.e. by setting $\delta V(t)=0$, and setting the bias voltage at $t=0$ to the constant value $V(t)=V_b>0$ under condition that the Josephson frequency $\Omega=2|e|V_b/\hbar$ is in resonance with mechanical frequency $\Omega=\omega$. Sign $\pm$ corresponds to terminal 1 and 2, respectively, i.e. to factor $-(-1)^l$ in the third, "$\varepsilon$"-term in Hamiltonian (\ref{Hamiltonian2CPB}). The momentum operator $\hat P = i(b^\dag - b)/\sqrt{2}$ acts upon the mechanical subsystem.
The operator (\ref{U_I}) is derived using the interaction picture, where the first two terms of Hamiltonian (\ref{Hamiltonian2CPB}), i.e. the "$\hbar\omega$"- and the "$E_J$"-term, are the noninteracting part $H_0$ leading to $\hat U_0(t,0)=\exp \left[ -\tfrac{i}{\hbar} \int_0^t H_0(t')\mathrm{d}t' \right] = \exp \left[ -i\tfrac{E_J}{\hbar \Omega}\sin \omega t - i \omega t b^\dag b \right]$, while the third, "$\varepsilon$"-term is the interacting one. In the interaction picture, this term appears as $\widetilde{H}_{int}=-\varepsilon E_J \left( \hat X \cos \omega t  + \hat P \sin \omega t \right) \left[ \cos \left( \tfrac{2E_J}{\hbar \Omega} \sin \Omega t \right) \sin \Omega t \, \sigma_y - \right. \\ \left. \sin \left( \tfrac{2E_J}{\hbar \Omega} \sin \Omega t \right) \sin \Omega t \, \sigma_z \right]$. Applying the resonant condition $\Omega = \omega$, using the periodicity of this Hamiltonian and rotating wave approximation, as well as condition $\varepsilon^2 N \ll 1$, under which one can approximately write $\hat U(NT,0) \approx \exp \left[ -\tfrac{i}{\hbar} N \int_0^T \widetilde{H}_{int}(t')\mathrm{d}t' \right]$, we finally come to expression (\ref{U_I}) \cite{npjDR}.\\

(II): The time-evolution operator
%
%%%%%%%%%%%%%%%%%%%%%%%%%%%%
\begin{eqnarray}\label{U_II}
\hat U_{\mathrm{II}}(\Delta t) = e^{-i \omega \Delta t\, b^\dag b} e^{-\frac{i}{\hbar} E_J \Delta t \cos \Phi \, \sigma_x }
\end{eqnarray}
%%%%%%%%%%%%%%%%%%%%%%%%%%%%
%
%
%%%%%%%%%%%%%%%%%%%%%%%%%%%%
%\begin{eqnarray}\label{U_II}
%\hat U_{\mathrm{II}}(\Delta t) &=& e^{-i \omega \Delta t\, b^\dag b} e^{-\frac{i}{\hbar} E_J \Delta t \cos \Phi^{(1)} \sigma_x^{(1)}} \nonumber \\
%&\times & e^{-\frac{i}{\hbar} E_J \Delta t \cos \Phi^{(2)} \sigma_x^{(2)}}
%\end{eqnarray}
%%%%%%%%%%%%%%%%%%%%%%%%%%%%
%
describes evolution of decoupled qubit from nanomechanics along the time interval $\Delta t$, under which mechanical states perform just gaining the phase factor $\exp(i\omega \Delta t)$ (unitary rotation in the phase space). It is obtained from Hamiltonian (\ref{Hamiltonian2CPB}) setting bias voltage and gate voltage to zero, i.e. $V(t)=\delta V(t)=0$, and superconducting phase $\Phi(t)$ constant. An additional request is that the time interval $\Delta t$ should be relatively small with respect to the time evolution interval of the system, i.e. $\Delta t \lesssim T$, so that the coupling term, i.e. the third term in Hamiltonian, $\varepsilon E_J \sim \sin \Phi$, can be neglected.\\

%superconducting phases should be set to $\Phi^{(1)}=\pi/2+2\pi n$, $\Phi^{(2)}=\pi/2+2\pi m$, $n,m$ are integers, and kept constant to eliminate action of the Josephson term, i.e. the second term in Hamiltonian $\sim \cos \Phi$.

(III): The time-evolution operator
%
%%%%%%%%%%%%%%%%%%%%%%%%%%%%
\begin{eqnarray}\label{U_III}
\hat U_{\mathrm{III}}\left( \tfrac{T}{2} \right)= e^{-i \pi \, b^\dag b} e^{\mp i \, \mathrm{sgn}(Z_{\pm})\eta \sigma_z},
\end{eqnarray}
%%%%%%%%%%%%%%%%%%%%%%%%%%%%
%
where $\eta=4|e|\mathcal{E} A/\hbar\omega$. In this form it is obtained for the time interval $\Delta t = T/2$ along which electric field $\mathcal{E} \neq 0$ exists, i.e. the gate voltage $\delta V(t) \neq 0$ is switched on, while the bias voltage $V(t)=0$ is switched off (SC phase is constant). Sign $\mp$ corresponds to terminal 1 and 2, respectively, i.e. to factor $-(-1)^l$ in the last, "$E_Z$"-term in Hamiltonian (\ref{Hamiltonian2CPB}). Also, due to relative shortness of $\Delta t$ and smallness of $\varepsilon$, the third term in Hamiltonian $\sim \varepsilon E_J \sin \Phi$ is neglected.
In the time-integration of this term, there appears $\int_0^{T/2} x_Z(t)\mathrm{d}t = A \int_0^{T/2} \cos (\omega t + \varphi_Z)\mathrm{d}t$. In our procedure, we set the time protocol exactly to provide two coherent states with $\varphi_+ = \pi/2$ and $\varphi_- = 3\pi/2$ as the initial state for this operator acting on them along a half-period (will be shown later). Result of integration is then $\int_0^{T/2} x_{Z\pm}(t)\mathrm{d}t = \mathrm{sgn}(Z_\pm)2A/\omega$, where $\mathrm{sgn}(Z_{\pm})=\mp 1$ corresponding to the sign of the half-plane of coordinate $x$ in the phase space where the particular coherent state propagates.
As stressed before, this form of operator is valid for well-developed coherent states.\\

Depending on applied control parameters in each terminal, by combining the corresponding operators $U_{\mathrm{I}} \, - \, U_{\mathrm{III}}$, we obtain the time-evolution operator for the network with two terminals, i.e. for the 2-qubit state.

\section{Transmission of the QI in the 2-qubit network}

\begin{figure}
\centerline{\includegraphics[width=\columnwidth]{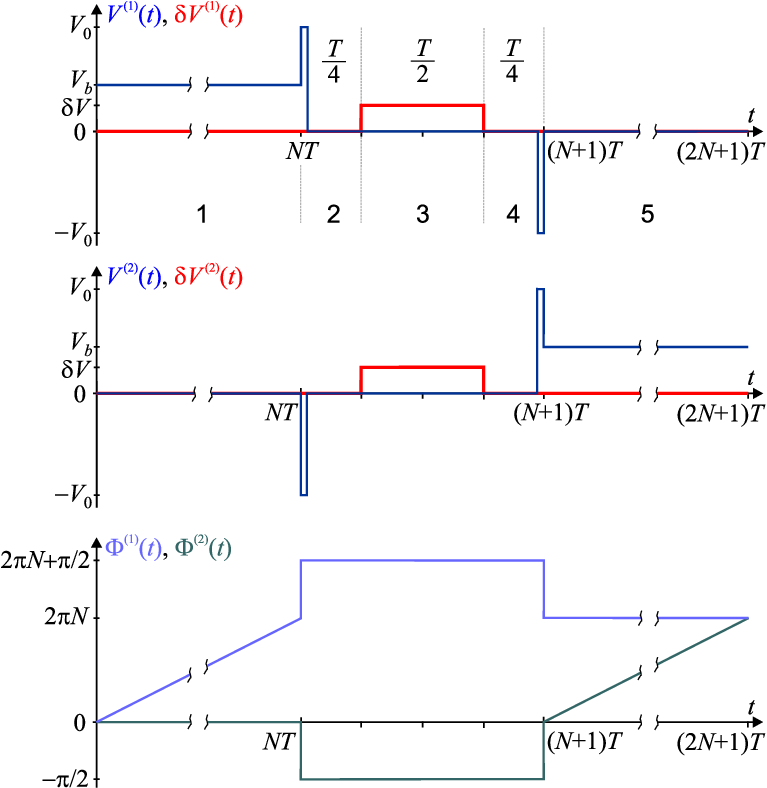}}
\caption{The time protocol of operating the control parameters of the 2-terminal network: bias voltage (blue), gate voltage (red) in terminal 1 (upper panel) and 2 (middle panel). The third panel shows the corresponding superconducting phase differences for Josephson junctions in terminal 1 (blue) and terminal 2 (green). Here, $N$ is an integer number and $T$ is period of mechanical oscillations. Numbers 1-5 denote five characteristic stages of time-evolution of the system.}
\label{Fig-TimeProtocol}
\end{figure}

\begin{figure*}
\centerline{\includegraphics[width=2\columnwidth]{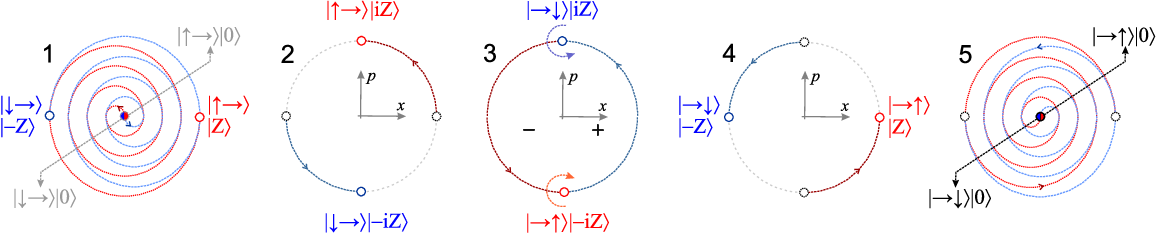}}
\caption{Schematic illustration of five characteristic stages of time evolution of the system in the $(x,p)$ phase space: (1) Josephson dynamics driven building of coherent states $\vert \pm Z \rangle$ entangled with qubit states $\vert \uparrow \rightarrow \rangle$, $\vert \downarrow \rightarrow \rangle$ from the pure state of Q1 (two colours are for two mechanical states that are generated); (2) unitary rotation of nanomechanical states by phase $\pi/2$; (3) electric filed driven  "rotation" of Q1 and Q2 states in which the QI is transmitted from Q1 states entangled with nanomechanics to Q2 states entangled with nanomechanics; (4) unitary rotation of nanomechanical states by phase $\pi/2$; (5) Josephson dynamics driven "deflating" of coherent states and creation of pure Q2 state.} 
\label{Fig-Transduction}
\end{figure*}

The operation of the 2-qubit network (Fig. \ref{Fig2qubit}) is governed by the time-evolution operators Eqs. (\ref{U_I}-\ref{U_III}). Each operator is activated/deactivated by properly setting control parameters (bias and gate voltages). The sequence of managing the control parameters we call the time-protocol. It activates the specific form of time-evolution operator, at the specific moment of time, to evolve the wave function of the system from that moment during the time interval of its operation $\Delta t$. The specific time-protocol for the desired goal of transmission of the QI between qubits is presented in Fig. \ref{Fig-TimeProtocol}, with the SC phases in terminals 1 and 2 during the time-evolution shown in the third panel of the figure. The time-protocol is divided in five sequential stages which are illustrated schematically in Fig. \ref{Fig-Transduction}.\\
%
%
%%%%%%%%%%%%%%%%%%%%%%%%%%
%%%%%%%%%%%%%%%%%%%%%%%%%%%%
%
%

%
%
%%%%%%%%%%%%%%%%%%%%%%%%%%
%%%%%%%%%%%%%%%%%%%%%%%%%%%%
%
%

\textbf{Stage 1} - building of nanomechanical coherent states and creating an entangled state with qubits (Fig. \ref{Fig-Transduction}-1):\\
This stage takes place during time interval $t \in (0,NT)$, i.e. $\delta t=NT$, with controlling parameters set to values: $V^{(1)}(t)=V_b>0$, $V^{(2)}(t)=0$, $\delta V^{(1)}(t)=\delta V^{(2)}(t)=0$. Time-evolution of Q1 is governed by $U_\mathrm{I}$ while Q2 undergoes $\hat U_\mathrm{II}$ type of time-evolution. The relevant operator is
%
%%%%%%%%%%%%%%%%%%%%%%%%%%%%
\begin{eqnarray}\label{U_1}
\hat U_1(NT,0)= e^{- i 2\pi N b^\dag b} e^{i \lambda N \hat P \sigma_y^{(1)}} e^{- i 2\pi \frac{E_J}{\hbar\omega} \, \sigma_x^{(2)}}.
\end{eqnarray}
%%%%%%%%%%%%%%%%%%%%%%%%%%%%
%
We obtain the wave function at the end of the stage 1 acting by $\hat U_1$ on the initial state (\ref{Psi_0}), i.e. $\Psi (NT) = \hat U_1(NT,0) \Psi(0)$. As the Q2 is in the eigenstate of $\sigma_x^{(2)}$ operator (third term in Eq. (\ref{U_1})), its state is not affected. Action on Q1, which is $\sim e_y^{\pm}$, appears through the second term in Eq. (\ref{U_1}) which can be written as $\cos(\lambda N \hat P)\, \mathbf{1} + i \sin(\lambda N \hat P)\, \sigma_y^{(2)}$. Each operator, $\mathbf{1}$ and $\sigma_y$, creates a corresponding state $e_y^-$ or $e_y^+$ of Q1, depending on its operand. These states are multiplied by $\cos(\lambda N \hat P)=(\exp(i\lambda N \hat P)+\exp(-i\lambda N \hat P))/2$ and $i\sin(\lambda N \hat P)=(\exp(i\lambda N \hat P)-\exp(-i\lambda N \hat P))/2$. Here, taking the coordinate representation of momentum operator $\hat P = -i x_0 \tfrac{\mathrm{d}}{\mathrm{d}x}$, we see that  
$\exp(\pm i \lambda N \hat P)=\exp(\pm \lambda N x_0 \tfrac{\mathrm{d}}{\mathrm{d}x})$ is a coordinate-shift operator by $\pm \lambda N x_0$, which by that create coherent states $\vert \pm Z \rangle \sim \exp \left[ -(x \mp \lambda N x_0)^2/2x_0^2  \right]$ acting on the mechanical ground state $\vert 0 \rangle \sim \exp \left[ -x^2/2x_0^2  \right]$. The first term in  Eq. (\ref{U_1})) generates the time-phase of mechanical states, i.e. $Z  \rightarrow Z \exp(-i2\pi N)$, which we denote as $\vert Z \rangle \equiv \vert Z \exp(-i2\pi N)  \rangle$, $\vert -Z \rangle \equiv \vert Z \exp(i2\pi N-i\pi)  \rangle$. Gathering all terms together, we finally get the wave function at $t=NT$ in the form
%
%%%%%%%%%%%%%%%%%%%%%%%%%%%%
\begin{eqnarray}\label{Psi_1}
\Psi (NT)= \alpha \vert \mathbf{e}_y^+\mathbf{e}_x^- \rangle \vert Z \rangle + \beta \vert \mathbf{e}_y^-\mathbf{e}_x^- \rangle \vert -Z \rangle.
\end{eqnarray}
%%%%%%%%%%%%%%%%%%%%%%%%%%%%
%

\textbf{Stage 2} - unitary rotation of entangled state by quater of period (Fig. \ref{Fig-Transduction}-2):\\
This stage takes place during time interval $t \in (NT,NT+T/4)$, i.e. $\Delta t=T/4$, with controlling parameters set to values: $V^{(1)}(t)=V^{(2)}(t)=0$, $\delta V^{(1)}(t)=\delta V^{(2)}(t)=0$. Time-evolution of both qubits is governed by $\hat U_\mathrm{II}$ type of time-evolution. At the beginning of the interval, at both terminals we apply short bias voltage pulse $V^{(1)}(t)=V^{(2)}(t)=V_0$ during very short time interval $\delta t \ll T$ to increase the SC phase difference in both terminals by $\pi/2$, i.e. $\Phi^{(1)}(t)=2\pi N + \pi/2$, $\Phi^{(2)}(t)=\pi/2$, keeping them constant during this period $T$. By this, we set the qubit-acting part of $\hat U_\mathrm{II}$ proportional to $\cos \Phi$ to zero and prevent perturbation of qubit states by Josephson dynamics during following one period of evolution. The small coupling term is neglected during this (short) interval.
The relevant operator is
%
%%%%%%%%%%%%%%%%%%%%%%%%%%%%
\begin{eqnarray}\label{U_2}
\hat U_2\left(NT+\tfrac{T}{4},NT\right)= e^{- i \frac{\pi}{2} b^\dag b},
\end{eqnarray}
%%%%%%%%%%%%%%%%%%%%%%%%%%%%
%
which just adds a phase factor $\exp (-i\pi/2)$ to nanomechanical states, i.e.
$\vert Z \rangle \rightarrow \vert Z \exp (-i2\pi N - i\pi/2) \rangle \equiv \vert i Z \rangle$, $\vert -Z \rangle \rightarrow \vert Z \exp (-i2\pi N - i3\pi/2) \rangle \equiv \vert -i Z \rangle$. The wave function at $t=NT+T/4$ is obtained by action of operator (\ref{U_2}) on state (\ref{Psi_1}) by which we finally get
%
%%%%%%%%%%%%%%%%%%%%%%%%%%%%
\begin{eqnarray}\label{Psi_2}
\Psi (NT+\tfrac{T}{4})= \alpha \vert \mathbf{e}_y^+\mathbf{e}_x^- \rangle \vert iZ \rangle + \beta \vert \mathbf{e}_y^-\mathbf{e}_x^- \rangle \vert -iZ \rangle.
\end{eqnarray}
%%%%%%%%%%%%%%%%%%%%%%%%%%%%
%

\textbf{Stage 3} - transmission of the QI between entangled nanomechanics and qubits (Fig. \ref{Fig-Transduction}-3):\\
This stage takes place during time interval $t \in (NT+T/4,NT+3T/4)$, i.e. $\Delta t=T/2$, with controlling parameters set to values: $V^{(1)}(t)=V^{(2)}(t)=0$, $\delta V^{(1)}(t)=\delta V^{(2)}(t)=\delta V$.
Dynamics in both terminals are governed by the $U_\mathrm{III}$ type of operator leading to
%
%%%%%%%%%%%%%%%%%%%%%%%%%%%%
\begin{eqnarray}\label{U_3}
\hat U_3\left(NT+\tfrac{3T}{4},NT+\tfrac{T}{4}\right)&=& e^{-i \pi \, b^\dag b} e^{-i \, \mathrm{sgn}(Z_{\pm})\eta_1 \sigma_z^{(1)}} \nonumber \\
&\times & e^{+i \, \mathrm{sgn}(Z_{\pm})\eta_2 \sigma_z^{(2)}}.
\end{eqnarray}
%%%%%%%%%%%%%%%%%%%%%%%%%%%%
%
Stage 1 of the evolution clearly defines the "well-developed coherent state" for which the $\hat U_\mathrm{III}$ (\ref{U_III}) and consequently $\hat U_3$ (\ref{U_3}) are valid: $\lambda N \gg 1$.
Using this operator we transmit the QI from Q1 to Q2 states entangled with nanomechanics in the following way: For Q1, by setting $\delta V^{(1)}$ (i.e. electric field contained in $\eta_1$) to fulfil the condition, imposed by the second term in (\ref{U_3}), $\exp \left[ \pm i \eta_1 \sigma_z^{(1)} \right] \, e_y^{\pm}\vert \pm i Z \rangle = \exp(\pm i \chi_1) \, e_x^- \vert \pm i Z \rangle$, we "rotate" both $e_y^{\pm}$ to the same state $e_x^-$. For that, it turns out that $\eta_1 = \pi/4$ and $\chi = -\pi/4$. Sign $\pm$ in the evolution operator comes from $\mathrm{sgn}(Z_{\pm})$ taking into account that operands, the mechanical states $\vert \pm i Z \rangle$, evolve along the negative/positive half-plane of $x$-coordinate, respectively. 
At the same time, for Q2 we set $\delta V^{(2)}$ (i.e. electric field contained in $\eta_2$) to fulfil the condition, imposed by the third term in (\ref{U_3}), $\exp \left[ \mp i \eta_2 \sigma_z^{(2)} \right] \, e_x^- \vert \pm i Z \rangle = \exp(\mp i \chi_2) \, e_y^{\pm} \vert \pm i Z \rangle$, i.e. to "rotate" the $e_x^-$ state of Q2 to states $e_y^{\pm}$.
For that, it turns out that $\eta_2 = \pi/4$ and $\chi_2 = -\pi/4$. It shows that $\eta_1=\eta_2$ and $\chi_1=\chi_2$, i.e. we apply the same gate voltage at both terminals as stated in the protocol. Furthermore, phase factors $\exp(\pm i \chi_{1,2})$, gained by "rotation" of Q1 and Q2, exactly cancel each other.
Finally, the first term in (\ref{U_3}) evolves mechanical states during a half-period adding a phase factor $\exp(-i\pi)$, i.e. $\vert i Z \rangle \equiv \vert Z \exp (-i2\pi N - i\pi/2) \rangle \rightarrow \vert Z \exp (-i2\pi N - i3\pi/2) \rangle \equiv \vert -i Z \rangle$, $\vert i Z \rangle \equiv \vert Z \exp (-i2\pi N - i3\pi/2) \rangle \rightarrow \vert Z \exp (-i2\pi N - i5\pi/2) \rangle \equiv \vert i Z \rangle$.
The wave function of the system at $t=NT+3T/4$ finally attains the form
%
%%%%%%%%%%%%%%%%%%%%%%%%%%%%
\begin{eqnarray}\label{Psi_3}
\Psi (NT+\tfrac{3T}{4})= \alpha \vert \mathbf{e}_x^- \mathbf{e}_y^+ \rangle \vert -iZ \rangle + \beta \vert \mathbf{e}_x^- \mathbf{e}_y^- \rangle \vert iZ \rangle.
\end{eqnarray}
%%%%%%%%%%%%%%%%%%%%%%%%%%%%
%

\textbf{Stage 4} - unitary rotation of entangled state by quater of period (Fig. \ref{Fig-Transduction}-4):\\
This stage takes place during time interval $t \in (NT+3T/4,(N+1)T)$, i.e. $\Delta t=T/4$.
It is entirely analogous to the stage 2, with the same controlling parameters, but here at the end of the evolution time interval we apply very short pulse of the bias voltage $-V_0$ in both terminals to decrease the SC phases by $-\pi /2$ (see Fig. \ref{Fig-TimeProtocol}, third panel). It is done in order to set the suitable initial conditions for the next stage of time-evolution.
The relevant operator is
%
%%%%%%%%%%%%%%%%%%%%%%%%%%%%
\begin{eqnarray}\label{U_4}
\hat U_4\left((N+1)T,NT+\tfrac{3T}{4}\right)= e^{- i \frac{\pi}{2} b^\dag b},
\end{eqnarray}
%%%%%%%%%%%%%%%%%%%%%%%%%%%%
%
and the wave function obtained by its action on (\ref{Psi_3}) is 
%
%%%%%%%%%%%%%%%%%%%%%%%%%%%%
\begin{eqnarray}\label{Psi_4}
\Psi ((N+1)T)= \alpha \vert \mathbf{e}_x^- \mathbf{e}_y^+ \rangle \vert Z \rangle + \beta \vert \mathbf{e}_x^- \mathbf{e}_y^- \rangle \vert -Z \rangle.
\end{eqnarray}
%%%%%%%%%%%%%%%%%%%%%%%%%%%%
%

\textbf{Stage 5} - "deflating" the nanomechanical coherent states and creating a pure qubit state (Fig. \ref{Fig-Transduction}-5):\\
This stage takes place during time interval $t \in ((N+1)T,(2N+1)T)$, i.e. $\delta t=NT$, with controlling parameters set to values: $V^{(1)}(t)=0$, $V^{(2)}(t)=V_b>0$, $\delta V^{(1)}(t)=\delta V^{(2)}(t)=0$. Time-evolution of Q1 is governed by $\hat U_\mathrm{II}$ while Q2 undergoes $\hat U_\mathrm{I}$ type of time-evolution. The relevat operator is
%
%%%%%%%%%%%%%%%%%%%%%%%%%%%%
\begin{eqnarray}\label{U_5}
\hspace{-1.cm} \hat U_5((2N + 1)T,(&N& + 1)T) = \nonumber\\
&& e^{- i 2\pi N b^\dag b} e^{- i 2\pi \frac{E_J}{\hbar\omega} \, \sigma_x^{(1)}} e^{-i \lambda N \hat P \sigma_y^{(2)}}.
\end{eqnarray}
%%%%%%%%%%%%%%%%%%%%%%%%%%%%
%
The SC phase in terminal 1 is constant, i.e. $\Phi^{(1)}(t)=NT$  (set to such value in previous stage), so the coupling term of Q1 to dynamics to nanomechanics ($\sim \sin \Phi$) is zero, thus preventing it to perturb the qubit state during this long period of evolution. As Q1 is now set to be in the eigenstate of $\sigma_x^{(1)}$, it is not affected by its action in (\ref{U_5}). Q2 undergoes dynamics completely analogous to the one of Q1 in stage 1. However, there is a substantial difference - the sign of the coupling term which generates a shift of nanomechanical coherent states is opposite. In that way, the same dynamics that lead to building ("inflating") of coherent states, now acts in opposite direction by reducing that shift and finally reducing ("deflating") both coherent states $\vert \pm Z \rangle$ to the nanomechanical ground state $\vert 0 \rangle$.
The wave function obtained at the end of stage 5 at $t=(2N+1)T$ is
%
%%%%%%%%%%%%%%%%%%%%%%%%%%%%
\begin{eqnarray}\label{Psi_5}
\Psi ((2N+1)T)= \left( \alpha \vert \mathbf{e}_x^- \mathbf{e}_y^+ \rangle + \beta \vert \mathbf{e}_x^- \mathbf{e}_y^- \rangle \right) \vert 0 \rangle.
\end{eqnarray}
%%%%%%%%%%%%%%%%%%%%%%%%%%%%
%
It is a pure state with the QI finally transferred from quantum superposition of states of Q1 into quantum superposition of states of Q2, by that fulfilling the given task of the 2-qubit network.

\section{Conclusions}

In this paper we propose a functional quantum network consisting of $M$ terminals, each containing a superconducting charge qubit placed between the voltage-biased superconducting leads in regime of the AC Josephson effect. Also, each terminal has a special setup of gate electrodes capable of providing a homogeneous electric field acting on each qubit at demand. Independent wiring of each terminal provides a mean to control it by two parameters, bias voltage and gate voltage, for which we can tailor a specific time-protocol of their operation to achieve a specific functionality of the network. All qubits in the network are attached to the same nanomechanical oscillator, i.e. a suspended beam oscillating in the $M$-th flexural mode. Coupled resonant dynamics of mechanically vibrating qubits with the Josephson tunneling between the leads and qubits results in formation of nanomechanical coherent states entangled with qubit states. As it is the same nanomechanical oscillator attached to all qubits, the mechanical state generated in each terminal affects them all - entanglement is spread along the whole network resulting in entangled multiqubit states and nanomechanical states. It is our main tool for transduction and transmission of quantum information in the network.\\

We illustrate in detail one possible functionality of the proposed network using example of 2-qubit network to transmit quantum information from one qubit to another. For that a specifically tailored time-protocol of operating (switching on/off to constant value or zero) the bias voltages and gate voltages in both terminals is proposed. Shortly, the quantum information is initially encoded in the pure state (quantum superposition of two states) of qubit-1. We apply a bias voltage in terminal-1 to utilize resonant Josephson dynamics coupled to mechanical oscillator to build the mechanical coherent states entangled with all qubit states, while keeping terminal-2 passive. In the next stage, applying the gate voltage in both terminals, we "rotate" the qubit states to the desired values so that now the quantum information is transmitted into superposition of entangled states of qubit-2. Finally, we apply the bias voltage in terminal-2, keeping terminal-1 passive, to reduce the entangled state of the system to the pure state with quantum information encoded in quantum superposition of the qubit-2 states.\\

The natural question that appears is about experimental feasibility of the proposed setup. It was mainly discussed in our first paper on this topic \cite{npjDR} performing number of numerical calculations simulating different decoherence and dephasing processes, mismatch from the resonance condition etc, as well as discussing with experimental experts regarding the fabrication, control of the applied external voltages and measurements. We came up with the list of parameters values, among which the charge-qubit decoherence time appears to be the bottleneck for the functionality. Having it of the order of nowadays achievable $1\mu$s, with mechanical quality factor $10^4$ or better, and bias voltages of the order of $10\mu$V controllable down to $0.1 \% $ (e.g. by Keysight B2961A), should satisfy the requirements. The fabrication of the very qubit is by coating the parts of the vibrating beam, e.g. a carbon nanotube, with aluminium to create islands of $10^{-22} - 10^{-21}$kg, large enough to sustain the superconductivity and to perform at $\sim 1$GHz of mechanical vibrations. The zero-point amplitude is then of the order of $1$pm, tunneling length of the order of 1\AA.\\

Besides transmission of quantum information between qubits facilitated by nanomechanics, the described procedures, controlled by the specific time-protocols, can give rise to other important tasks in the quantum information processing. Those are, for example, codes for quantum error correction having the no-cloning theorem \cite{Park} in mind: the 3-qubit bit flip code \cite{Peres}, or the 9-qubit Shor code \cite{Shor}. Nanomechanical implementation of such, together with challenges emerging from it, is a subject of our future research.\\

\section*{Acknowledgements}

This work was supported by the QuantiXLie Centre of Excellence, a project co-financed by the Croatian
Government and European Union through the European Regional Development Fund - the Competitiveness and Cohesion Operational Programme (Grant KK.01.1.1.01.0004), and by IBS-R024-D1. The authors are grateful to the PCS IBS, Daejeon, Republic of Korea for the hospitality while working on this paper.

\end{document}